\documentclass[%
 reprint,
 amsmath,amssymb,
 aps,
 prl,
]{revtex4-1}

\usepackage[dvips]{graphicx}% Include figure files
\usepackage{dcolumn}% Align table columns on decimal point
\usepackage{bm}% bold math
\usepackage{xcolor}
\definecolor{lgtblue}{RGB}{0,128,255}
\definecolor{spinach}{RGB}{0,193,0}
\definecolor{orng}{RGB}{255,128,0}
\definecolor{pnk}{RGB}{255,74,165}
\definecolor{darkblue}{RGB}{0,0,160}

\begin{document}

\preprint{APS/123-QED}

\title{Coupling of elasticity to capillarity in soft aerated materials}% Force line breaks with \\
\thanks{Financial support from Saint-Gobain Recherche (Aubervilliers, France) is acknowledged.}%

\author{Lucie Duclou\'e}
\email{lucie.ducloue@ifsttar.fr}%
\author{Olivier Pitois}%
\author{Julie Goyon}%
\author{Xavier Chateau}%
\author{Guillaume Ovarlez}%
\affiliation{%
 Laboratoire Navier (UMR CNRS 8205), Universit\'e Paris-Est, Champs-sur-Marne, France
}%

\date{\today}% It is always \today, today,
             %  but any date may be explicitly specified

\begin{abstract}
We study the elastic properties of soft solids containing air bubbles. Contrary to standard porous materials, the softness of the matrix allows for a coupling of the matrix elasticity to surface tension forces brought in by the bubbles. Thanks to appropriate experiments on model systems, we show how the elastic response of the dispersions is governed by two dimensionless parameters: the gas volume fraction and a capillary number comparing the elasticity of the matrix to the stiffness of the bubbles. We also show that our experimental results are in good agreement with computations of the shear modulus through a micro-mechanical approach.
%\begin{description}
%\item[Usage]
%Secondary publications and information retrieval purposes.
%\item[PACS numbers]
%May be entered using the \verb+\pacs{#1}+ command.
%\item[Structure]
%You may use the \texttt{description} environment to structure your abstract;
%use the optional argument of the \verb+\item+ command to give the category of each item. 
%\end{description}
\end{abstract}

%\pacs{Valid PACS appear here}% PACS, the Physics and Astronomy
                             % Classification Scheme.
%\keywords{Suggested keywords}%Use showkeys class option if keyword
                              %display desired
\maketitle

\section*{Introduction}

Complex systems of a dispersed phase in a solid matrix can behave very differently from one of their components taken alone. Their broad range of properties explains that examples of dispersions such as composites~\cite{matthews1999composite} or porous media~\cite{coussy2011mechanics} are widespread in nature and in the industry. In all dispersions, interfacial forces can appear at the boundary between the dispersed phase and the continuous matrix. A coupling of surface tension forces to the bulk elasticity of a solid has been evidenced in soft systems like biological tissues~\cite{clements1961pulmonary}, or through the deformation of soft substrates like polymers at the contact line with a drop resting on the solid~\cite{pericet2008effect}. Capillary forces also affect the overall mechanical properties of nanoporous media~\cite{duan2005size}. For larger pores, because of the hardness of the matrix in usual porous media, the influence of interfacial effects on the overall properties of the saturated material is negligible~\cite{dormieux2006microporomechanics}. Dispersions in softer materials could allow for observable coupling of interfacial forces to the bulk elasticity of the solid at larger scales than the nanometer. Many dense suspensions~\cite{coussot2005rheometry} of geological interest, like muds, or with industrial applications, like fresh concrete or emulsions, behave as soft elastic solids below a critical level of stress~\cite{coussot2012rheophysique}. To study the role of surface tension forces in soft elastic materials, we investigate the elastic behaviour of dispersions of bubbles in concentrated emulsions. Those aerated emulsions, which have applications in the food~\cite{vanAken2001333} and cosmetic~\cite{balzer1991alkylpolyglucosides} industry, have been the subject of stability and rheology studies~\cite{C1SM06537H, PhysRevLett.104.128301, kogan2013mixtures}. However, their overall elastic properties have not yet been studied in detail.

In dispersions of bubbles in a soft material, coupling between the elasticity of the matrix and capillary effects is expected to occur through bubble deformation. The elastic deformation of the matrix tends to deform the bubbles and surface tension forces will thus act to minimize the area of the bubble by maintaining a spherical shape. The limit case of negligible surface tension forces is a soft porous medium. Theoretical work shows that adding holes in a solid softens it~\cite{dormieux2006microporomechanics}. In the limit case of predominant surface tension forces compared to the matrix elasticity, a bubble should no longer be deformable and should behave as a rigid inclusion with no shear stiffness. Experimental and theoretical work have shown that rigid beads in a soft solid strengthen the solid~\cite{mahaut2008yield}. The case of rigid bubbles is similar except for the boundary condition, changed from no-slip for beads to full-slip for bubbles. Theoretical models in the dilute limit predict a strengthening of the dispersion when adding rigid bubbles~\cite{dormieux2006microporomechanics}. Between those two limit cases, more work is needed to investigate the elastic response of the soft aerated solid. In this work, we restrain to the range of gas volume fraction $\phi<50\%$, so that we do not consider foams of those materials, in which the bubbles are deformed by geometrical constraints. We design model systems and appropriate experimental methods that allow us to measure the shear modulus of dispersions of monodisperse bubbles embedded in a medium of chosen elasticity. We compare our experimental results to estimates of the elastic modulus through a micro-mechanical approach. 
%pas mis importance de ce résultat
\section*{Experimental aspects}
\begin{figure}
\centering
\includegraphics[scale=0.6]{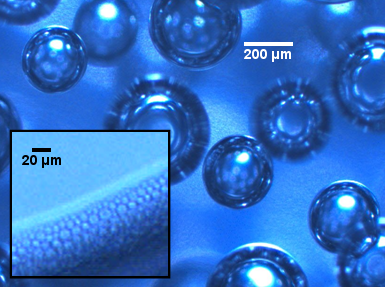}
\caption{Microphotograph of a dispersion of monodisperse bubbles (R=200$\mu$m) in emulsion (2). The emulsion is transparent, allowing for the visualisation of in-depth bubbles that thus do not have the same apparent radius. Inset: close-up of droplets of emulsion (2) at the interface with a bubble.\label{fig:photo}}
\end{figure}

The dispersion matrices we choose are concentrated oil in water emulsions of shear moduli ranging from 100 to 1000Pa. Concentrated emulsions behave as soft elastic solids for stresses well below their yield stress~\cite{mason1995elasticity}. In the experimental systems, unless otherwise indicated, the radius of the droplets is around 1 to 2$\mathrm{\mu m}$ (the poydispersity is around 20\%), which, at the considered gas volume fractions, should ensure that there is scale separation between the drops and the bubbles, and consequently validate the use of the emulsion as an elastic continuous medium embedding the bubbles~\cite{goyon2008spatial}. In all the systems, the yield stress of the emulsion is high enough to ensure that no bubble rise occurs at rest or during measurements~\cite{Dubash2007123}. Most dispersions are prepared by gently mixing the emulsion with a separately produced monodisperse foam. The foams are obtained by blowing nitrogen plus a small amount of perfluorohexane ($\mathrm{C_{6}F_{14}}$) through a porous glass frit or through needles: we are able to produce nearly monodisperse foams with average bubble radii $R_{b}$ ranging from 40$\mathrm{\mu m}$ to 800$\mathrm{\mu m}$. Coarsening is strongly reduced by the presence of $\mathrm{C_{6}F_{14}}$~\cite{gandolfo1997interbubble}, meaning that the bubble size is stable during measurements. The continuous phase of the foam is the same as the one in the emulsion, ensuring that mixing is easy and does not induce any chemical effect in the dispersions. The mixing with the foam adds a small amount of continuous phase to the emulsion. To ensure that for a series of experiments at different $\phi$ in a given emulsion, the elastic modulus of the matrix in the dispersions remains the same, we add controlled amounts of pure continuous phase in order to reach the same the oil volume fraction in the emulsion~\cite{PhysRevLett.104.128301, kogan2013mixtures}. An example of a dispersion of bubbles in an emulsion is shown on figure~\ref{fig:photo}. The composition of all the tested emulsions is indicated in table~\ref{tab:recap_systs}, and illustrates the variety of chemical compositions, surface tensions and elastic properties of the matrix that were used to perform the study.

The shear modulus of the dispersions is measured on a control stress rheometer by imposing small amplitude oscillations at a frequency of typically 1Hz. The oscillatory stress is chosen to be well below the yield stress of the systems, so that the oscillations are performed in the linear elastic regime of each material. At this frequency, the loss modulus of the systems is negligible. The geometry used to perform the rheometrical measurements is chosen according to the bubble size : for $R_{b}\le 50\mathrm{\mu m}$, the material is sheared between parallel  plates (radius $R$=25mm, gap $h$=2.5mm). The planes are serrated to prevent slippage of the dispersion~\cite{coussot2005rheometry}. Dispersions containing bigger bubbles require a larger thickness of sheared material and are studied in Couette-like devices : for $50\mathrm{\mu m}< R_{b}< 800\mathrm{\mu m}$, we use a vane in cup (exceptionally a serrated bob in cup) geometry (inner radius $R_i$=12.5mm, outer radius $R_o$=18mm), and for $R_{b}\ge 800\mathrm{\mu m}$, we use vane in cup geometries (either $R_i$=12.5mm and $R_o$=25mm or $R_i$=22.5mm and $R_o$=45mm). 
\begin{center}
\begin{table*}
{\renewcommand{\arraystretch}{1.5}
\renewcommand{\tabcolsep}{0.2cm}
\begin{tabular}{|l|c|c|c|c|}
\hline
& \textbf{oil - vol. fraction} & \textbf{continuous phase} & \textbf{$\mathbf{G'(0)}$ (Pa)} & \textbf{$\mathbf{\gamma}$ (mN.$\mathbf{m^{-1}}$)}\\
\hline
emulsion (1a) & silicon (V20) - 75\% & Forafac$\textregistered$ ($\mathrm{{DuPont}^{TM}}$) 4\% w. in water & 230 &15.5 $\pm$ 0.1 \\
\hline
emulsion (1b) & silicon (V20) - 73\% & Forafac$\textregistered$ ($\mathrm{{DuPont}^{TM}}$) 4\% w. in water & 163 &15.5 $\pm$ 0.1 \\
\hline
emulsion (2) & silicon (V350) - 79\% & TTAB 3\% w. in water/glycerol 50/50 w/w & 650 & 35.5 $\pm$ 0.1 \\
\hline
emulsion (3) & dodecane - 73\% & SDS 2.7\% w. in water & 285 & 36 $\pm$ 1 \\
\hline
emulsion (4) & silicon (V350) - 70\% & TTAB 3\% w. in water/glycerol 36/64 w/w & 799 & 35 $\pm$ 1 \\
\hline
\end{tabular}}
\caption{Synthetic description of all the emulsions used as matrices in the bubble dispersions: nature and volume fraction of the oil dispersed phase, composition of the aqueous continuous phase (including the surfactant) and relevant physical constants for the determination of the capillary number: elastic modulus of the matrix, and surface tension between air and the continuous phase. The composition given is the one of the matrix actually embedding the bubbles.\label{tab:recap_systs}}
\end{table*}
\end{center}
\section*{Results}
\begin{figure}
\centering
\includegraphics[scale=0.29]{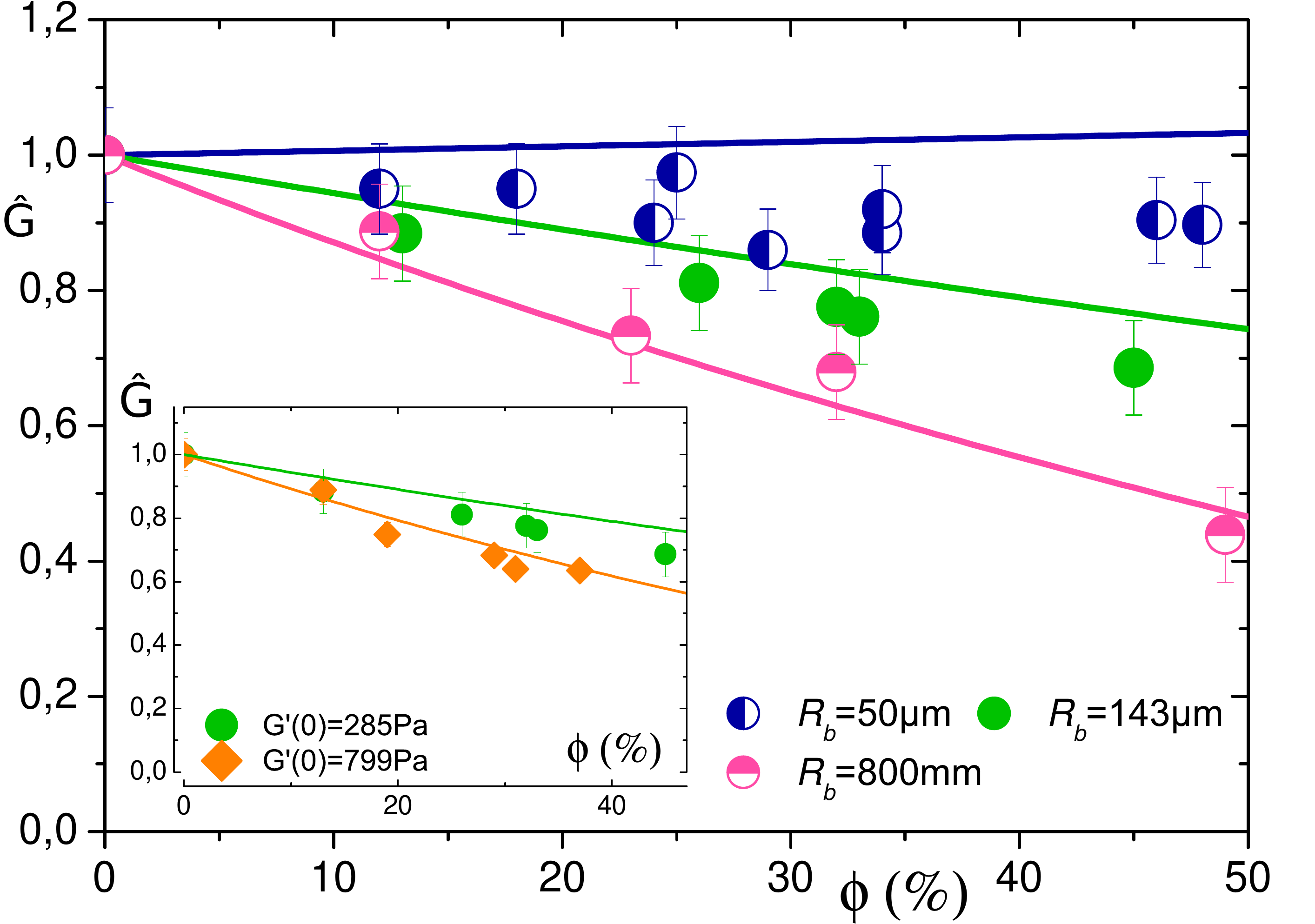}
\caption{Dimensionless elastic modulus $\hat{G}$ as a function of the gas volume fraction $\phi$ for dispersions with three different bubble radii $R_b$ in emulsion (3) [see legend]. The full lines are the computed $\hat{G}_{homog}(\phi)$ for $Ca=0.23$ (dark blue), $Ca=0.57$ (green) and $Ca=3.2$ (pink); experimentally measured $Ca$: $0.23\pm0.05$, $0.57\pm0.08$, $3.2\pm0.4$. Inset: $\hat{G}$ as a function of $\phi$ for dispersions of $R_b\approx 150\mathrm{\mu m}$ in \textcolor{spinach}{$\bullet$} emulsion (3) and \textcolor{orange}{$\blacklozenge$} emulsion (4). The full lines are the computed $\hat{G}_{homog}(\phi)$ for $Ca=0.57$ (green) and $Ca=1.65$ (orange); experimentally measured $Ca$: $0.57\pm0.08$, $1.65\pm0.15$.\label{fig:1} }
\end{figure}

\begin{figure}
\centering
\includegraphics[scale=0.30]{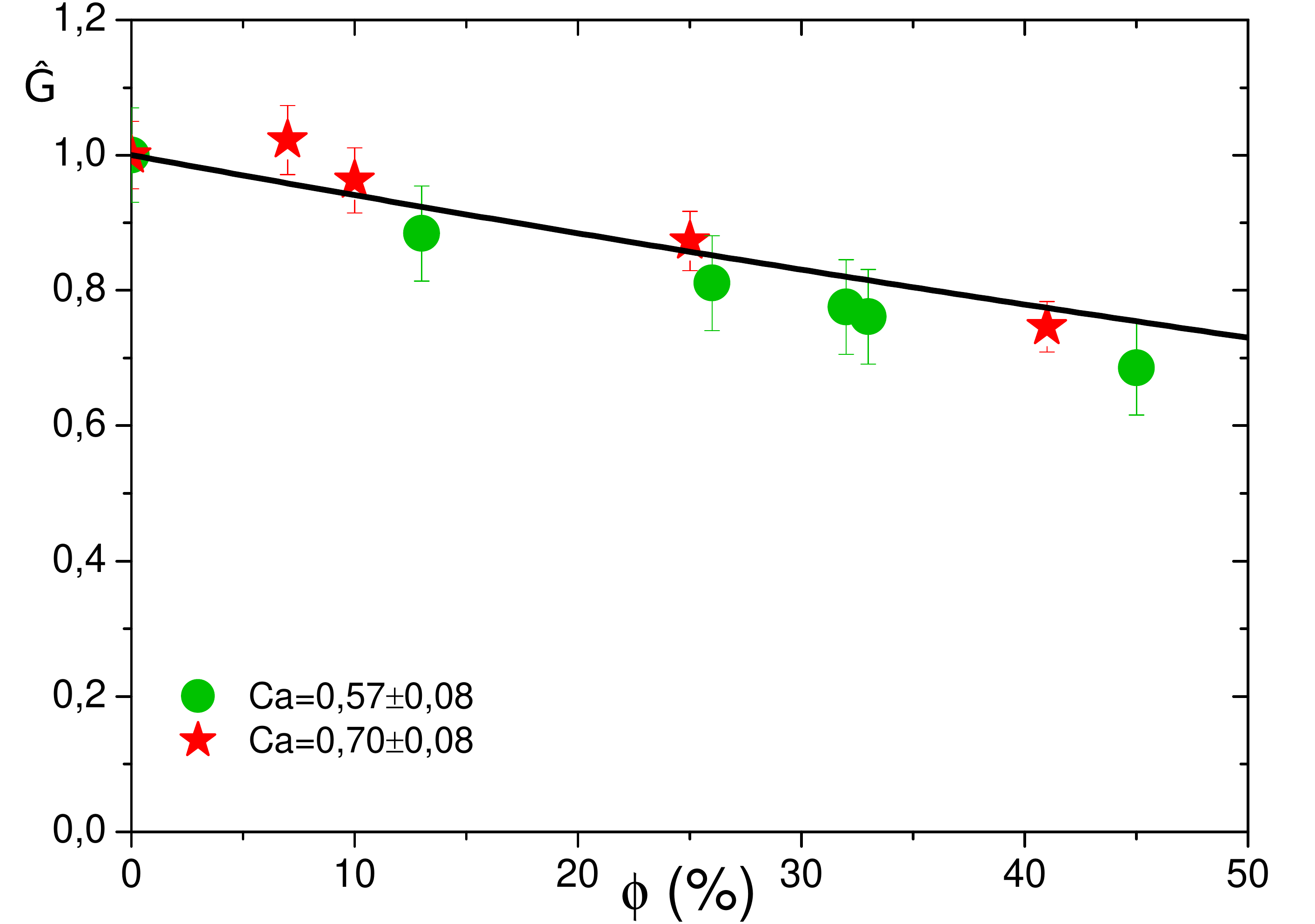}
\caption{Effect of a change in the surface tension: dimensionless elastic modulus $\hat{G}$ as a function of $\phi$ for dispersions of \textcolor{spinach}{$\bullet$} $R_b=143\mathrm{\mu m}$ bubbles in emulsion (3) and \textcolor{red}{$\bigstar$} $R_b=129\mathrm{\mu m}$ bubbles in emulsion (1). The surface tension is much lower in emulsion (1b), but $G'(0)$ has been chosen to get close values for $Ca$ in both systems. This experimentally measured $Ca$ are $0.57\pm0.08$ and $0.70\pm0.08$. The full line is the computed $\hat{G}_{homog}(\phi, Ca)$ at $Ca=0.63$, which is compatible with both systems, given the uncertainty on the value of $Ca$.\label{fig:2}}
\end{figure}

We start by studying the influence of the bubble radius $R_b$. In this aim, we prepare dispersions of bubbles in emulsion (3) (see table~\ref{tab:recap_systs} for details). In a first series of experiments, we add bubbles of $R_b=(50\pm10)\mathrm{\mu m}$ ($10\mathrm{\mu m}$ being the width of the volume-weighed bubble radius distribution) at various gas volume fractions $\phi$ in the emulsion. Those bubbles are slightly more polydisperse than is generally used for this study, because of the foam production technique. The shear modulus $G'(\phi)$ of the dispersions is measured to be slightly decreasing with $\phi$. This result is reported in dimensionless quantities $\hat{G}(\phi)=G'(\phi)/G'(0)$ as a function of $\phi$ on figure~\ref{fig:1}. We then prepare dispersions of larger bubbles in emulsion (3): a series with $R_b=(143\pm17)\mathrm{\mu m}$ and another one with $R_b=(800\pm40)\mathrm{\mu m}$. The results for $\hat{G}(\phi)$ are also reported on figure~\ref{fig:1}. The measurements show that the larger the bubbles, the softer the dispersion. This result can be understood as a manifestation of a simple physical effect, already evidenced in~\cite{kogan2013mixtures} (see also~\cite{rust2002effects} and~\cite{llewellin2002rheology} for the effect of bubble deformation on the viscosity of bubbly Newtonian fluids), that the interfacial energy to volume ratio is lower in larger bubbles, resulting in least bubble resistance to deformation. 
We now keep the bubble size constant, and vary the elastic modulus of the matrix: we prepare dispersions of $R_b=143\mathrm{\mu m}$ bubbles in emulsion (3) and of $R_b=(150\pm10)\mathrm{\mu m}$ bubbles in emulsion (4) (see table~\ref{tab:recap_systs}). In the two series of experiments, the bubble sizes are close and the surface tension is similar, but $G'(0)$ is almost three times higher in emulsion (4). $\hat{G}(\phi)$ is plotted for both systems on the inset in figure~\ref{fig:1}. As observed on the previous suspensions, $\hat{G}(\phi)$ is a decreasing function of $\phi$, and this decrease is all the stronger as $G'(0)$ is high.
To quantify the competition between the matrix elasticity and the bubble's resistance to deformation, we introduce a capillary number 
\begin{equation}
Ca=\frac{G'(0)}{2\gamma/R_b}
\label{eq:Ca}
\end{equation} 
which compares the shear modulus of the dispersion medium to the interfacial stress scale, the capillary pressure in the bubbles. This capillary number is equally affected by an increase in $R_b$ or a decrease in $G'(0)$. To quantify the relevance of $Ca$ on the overall elastic response of the dispersion at a given $\phi$, we perform two series of experiments with close $R_b$, but very different capillary pressure because of very different surface tension, and we adjust the elastic modulus in one of the emulsions so that $Ca$ is similar in both systems. The two experimental systems are as follow: the first one is dispersions of $R_b=143\mathrm{\mu m}$ of radius bubbles in emulsion (3), which leads to $Ca=0.57\pm0.08$, and the second one dispersions of (129$\pm$10)$\mathrm{\mu m}$ of radius bubbles in emulsion (1b) for which $Ca=0.70\pm0.08$. We observe that the measured values of $\hat{G}(\phi, Ca)$ are very close, as can be seen on figure~\ref{fig:2}. The value of $Ca$ unequivocally determines the elastic behaviour of the dispersion at a given $\phi$.

\begin{figure}
\centering
\includegraphics[scale=0.30]{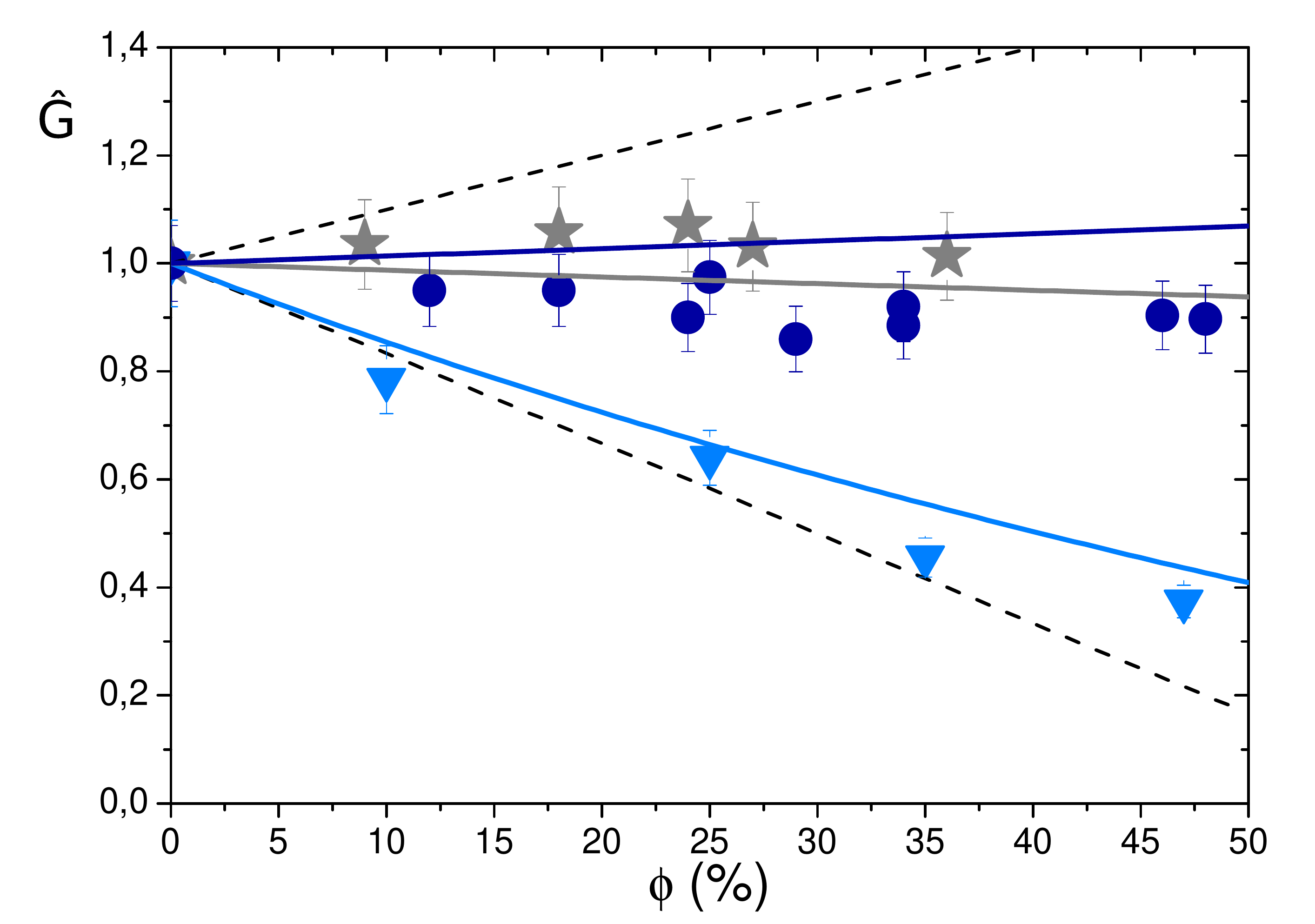}
\caption{Remarkable values of $Ca$: $\mathbf{Ca\to \infty}$: $\hat{G}(\phi)$ for dispersions of \textcolor{lgtblue}{$\blacktriangledown$} $R_b$=1mm bubbles in emulsion (2). The full line is the computed $\hat{G}_{homog}(\phi, Ca)$ at $Ca=9.0$ (light blue); experimentally measured $Ca$: $9.0\pm1.2$. $\mathbf{Ca\approx 0.25}$: dimensionless elastic modulus $\hat{G}$ for dispersions of \textcolor{darkblue}{$\bullet$} $R_b$=50$\mu$m bubbles in emulsion (3) and \textcolor{gray}{$\bigstar$} $R_b$=41$\mu$m in emulsion (1a). The full lines are the computed $\hat{G}_{homog}(\phi, Ca)$ at $Ca=0.23$ (dark blue) and $Ca=0.30$ (grey); experimentally measured $Ca$: $0.23\pm0.05$, $0.30\pm0.05$. The dashed lines are the dilute limits for rigid (top) and fully deformable (bottom) spheres, with a full slip boundary condition.\label{fig:3} }
\end{figure}

We now investigate the limit value of $Ca\to \infty$, for which surface tension forces are negligible and the bubbles can be assimilated to holes in the matrix. This is the case in usual porous materials. $\hat{G}(\phi, Ca\to\infty)$ can then be computed in the dilute limit~\cite{dormieux2006microporomechanics}: $\hat{G}(\phi, Ca\to\infty)=1-\frac{5}{3}\phi$. To compare this prediction to experimental data, we design a system in which surface tension effects are bound to be poor: we include the biggest bubbles of this study, of radius (1 $\pm$ 0.1) mm, in emulsion (2), which has a high elastic modulus (see table~\ref{tab:recap_systs}). Note that for this system the bubbles are injected directly in the emulsion in a tee-junction in a milli-fluidic device. As before, we measure the elastic modulus of the dispersion at various $\phi$. The experimental data points for the dimensionless modulus $\hat{G}(\phi)$ are compared to the dilute limit for dispersions of holes in an elastic medium on figure~\ref{fig:3}. We observe that $\hat{G}$ is a decreasing function of $\phi$. The exact value of $Ca$ in this system is $9.0\pm1.2$. The theoretical dilute limit for spherical holes in an elastic medium is already a good estimate of $\hat{G}(\phi\to 0, Ca)$at $Ca\sim10$.

\begin{figure}
\centering
\includegraphics[scale=0.3]{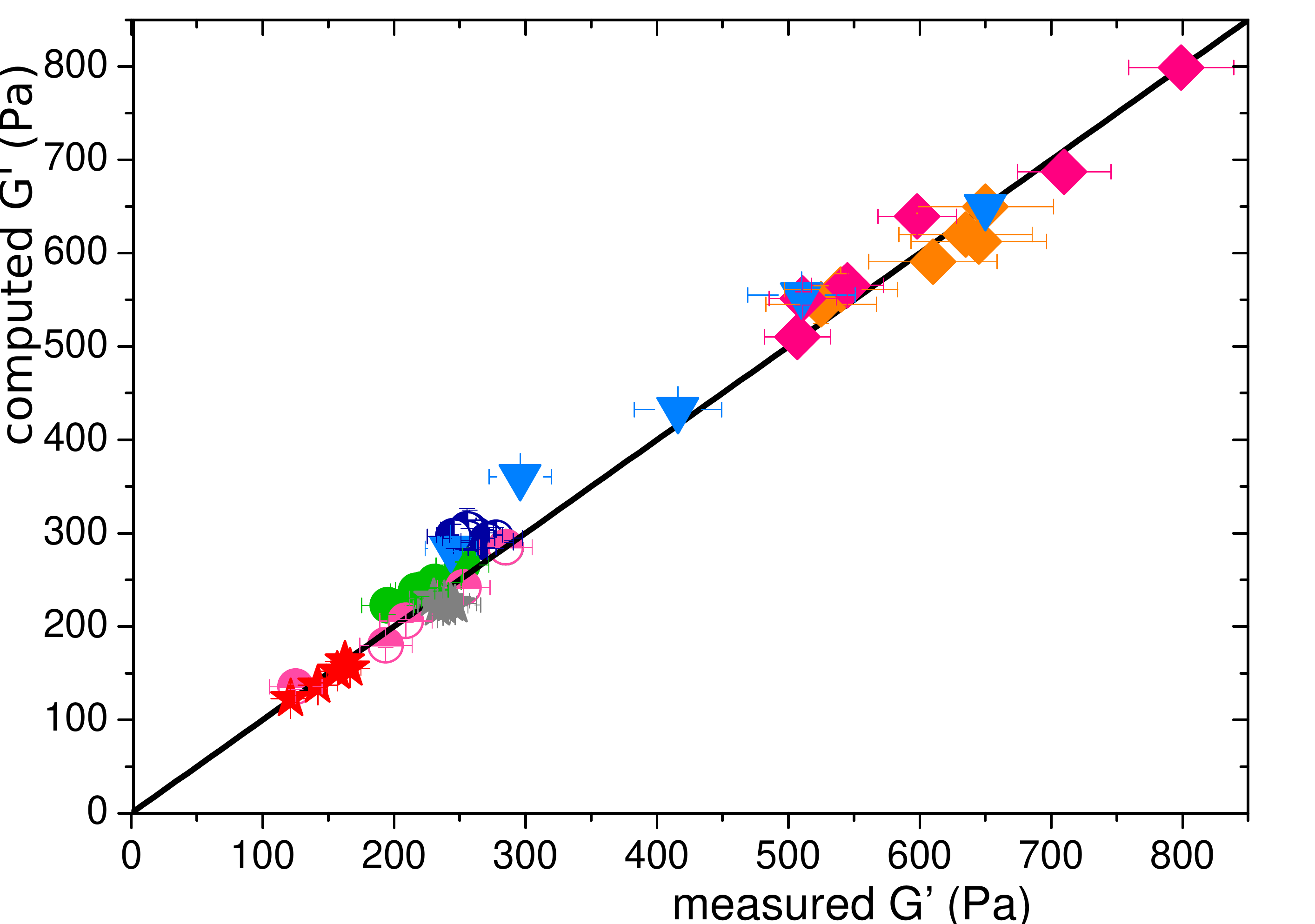}
\caption{Consistency between $G'$ and ${G'}_{homog}$ for all tested systems. $\bigstar$: emulsion (1), $\blacktriangledown$: emulsion (2), $\bullet$: emulsion (3), $\blacklozenge$: emulsion (4).\label{fig:4} }
\end{figure}

The limit case of $Ca\to 0$ also leads to a simplification: the bubbles are stiff compared to the matrix and the dispersion is made of rigid spheres with a full slip boundary condition in an elastic medium. The theoretical dilute limit can be computed as $\hat{G}(\phi, Ca=0)=1+\phi$~\cite{dormieux2006microporomechanics} and is plotted on figure~\ref{fig:3}. An experimental validation of this limit with our systems may be biased, because increasing the capillary pressure would mean reducing $R_b$, and we might no longer assume scale separation between the bubbles and the oil droplets. We thus choose not to investigate this limit.
From $Ca\to 0$ to $Ca\to \infty$, $\hat{G}(\phi, Ca)$ turns from an increasing to a decreasing function of $\phi$. Between these two extreme values, we have observed on the dispersion of the smallest bubbles in emulsion (3), already discussed above on figure~\ref{fig:1}, that $G'(\phi)$ has little variation with $\phi$ and is comparable to $G'(0)$. The capillary number in this system is $Ca=0.23\pm0.05$. To further check the peculiarity of this value of $Ca$, we prepare another dispersion of small bubbles $R_b$=(41$\pm$5)$\mathrm{\mu m}$ in emulsion (1a) (see table~\ref{tab:recap_systs}), with a close capillary number: $Ca=0.30\pm0.05$. $\hat{G}(\phi, Ca)$ for both dispersions of small bubbles is plotted on figure~\ref{fig:3}. We observe that in both systems, $\hat{G}(\phi, Ca)$ exhibit little dependence on the gas volume fraction, and is of order 1. The non-perturbative effect of bubble addition in the matrix can be seen as an experimental validation of previous micro-mechanical calculations~\cite{palierne1991rheologica, doi:10.1061/9780784412992.224} which have shown that a spherical bubble of radius $R_b$ and surface tension $\gamma$ in an elastic medium can be described as an equivalent elastic sphere of radius $R_b$ and no surface tension. Indeed, the deformation of the bubble under a strain $\epsilon$ leads to an increase in the bubble area that is proportional to $\epsilon^2$. The stored interfacial energy scales as $\gamma \epsilon^2$, which is analogous to an elastic energy. If the equivalent elasticity of the sphere is equal to that of the matrix, the bubbles are non-perturbative and $\hat{G}(\phi)= 1$. The equivalent elasticity of a bubble in a matrix $G'(0)$ can be written as a function of $G'(0)$ and $Ca$~\cite{palierne1991rheologica, doi:10.1061/9780784412992.224}:
\begin{equation}
G^{eq}=G'(0)\frac{8}{3+20Ca}
\end{equation}
with $Ca$ defined in equation~\ref{eq:Ca}. The expression of $G^{eq}$ shows that $Ca$ introduced above does not actually compare the equivalent elasticity of the bubble to that of the matrix. This explains why the overall elasticity of the dispersion is unperturbed by the presence of the bubbles for a somewhat unnatural value of $Ca$ around 0.2 to 0.3, which can be understood thanks to the computation of $G^{eq}$: $G^{eq}=G'(0)$ for $Ca=1/4$. Relying on the equivalent elastic sphere model for a bubble, a micro-mechanical approach allows to compute the overall elastic properties of the dispersions at finite $Ca$. The overall elasticity of a composite material made of elastic spheres in a matrix of another elastic material in the semi-dilute limit can be computed as a function of $Ca$ and $\phi$, in the framework of the Mori-Tanaka scheme~\cite{doi:10.1061/9780784412992.224, palierne1991rheologica}: 
\begin{equation}
\hat{G}_{homog}(\phi, Ca)=1-\frac{\phi(4Ca-1)}{1+\frac{12}{5}Ca-\frac{2}{5}\phi(1-4Ca)}
\end{equation}
Note that this expression is compatible with the previously discussed limits of $Ca \to \infty$, $Ca \to 0$ and $Ca=1/4$. Predictions of the model for $\hat{G}(\phi, Ca)$ at the experimentally measured $Ca$ are plotted in full coloured lines on figures~\ref{fig:1} to~\ref{fig:3}. A comparison of $\hat{G}_{homog}(\phi, Ca)$ to $\hat{G}(\phi, Ca)$ for all the systems we used, at all tested gas volume fractions is presented on figure~\ref{fig:4}. Experimental measurements and computations are generally in good agreement all over the range of systems we investigated. 

\begin{figure}
\centering
\includegraphics[scale=0.32]{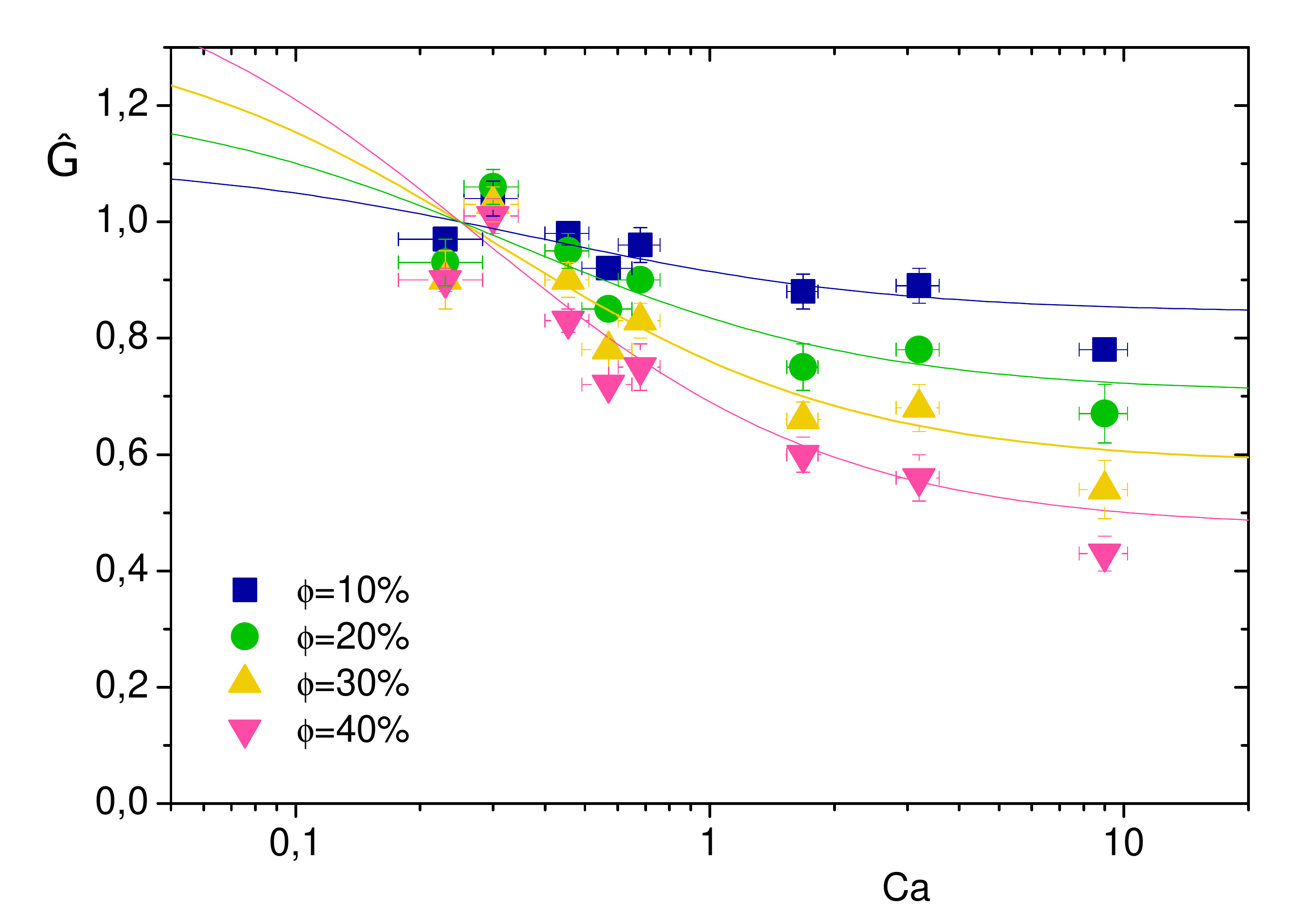}
\caption{Dimensionless elastic modulus $\hat{G}$ as a function of $Ca$ for four different values of the gas volume fraction. The dots are interpolated experimental data points, the full lines are $\hat{G}_{homog}(\phi, Ca)$. \label{fig:5} }
\end{figure}

As we have seen that the two dimensionless parameters $Ca$ and $\phi$ are enough to understand and predict the elasticity of the dispersions, we now plot $\hat{G}(\phi, Ca)$ as a function of $Ca$, for 4 values of $\phi$, on figure~\ref{fig:5}. As can be noticed on the graphs~\ref{fig:1} to~\ref{fig:3}, the achieved values of $\phi$ are different for all tested systems. To be able to plot $\hat{G}(\phi,Ca)$ at a given $\phi$, we interpolate the experimental data at the exact values of $\phi$ used for plotting on figure~\ref{fig:5}. The full lines are computations of $\hat{G}_{homog}(\phi, Ca)$. As expected, $\hat{G}(\phi, Ca)$ is a decreasing function of $Ca$: higher values of $Ca$ correspond to more deformable bubbles that lower the overall elastic modulus of the dispersions. The non-perturbative effect of the bubbles for $Ca=1/4$ is evidenced by the crossing of $\hat{G}_{homog}(\phi, Ca)$ at 1 for $Ca=0.25$, whatever the gas volume fraction. Below this value, the increase of $\hat{G}_{homog}(\phi, Ca)$ is consistent with previously discussed theoretical limits, but could not be investigated with our experimental systems. The series of data points at $Ca=0.23\pm0.05$ does not fit in the increasing $\hat{G}_{homog}(\phi, Ca)$ regime, perhaps because of broader polydispersity: the uncertainty on the value of $Ca$ mainly arises from the width of the bubble radius distribution and the value of $Ca$ for the largest bubbles in the foam is for instance higher than 0.25. A model computing $\hat{G}_{homog}(\phi, Ca)$ as a function of the whole measured distribution of radii may better represent the experimental data.

\section*{Conclusions}
As a conclusion, we have designed model systems in which precise control of the bubble stiffness and the matrix elasticity allows to experimentally determine the elastic modulus of dispersions of bubbles in a soft matrix. The results show that the addition of bubbles leads to a softening of the dispersion that is finely tuned by the capillary number. Those model systems enable us to compare our experimental results to estimates of the shear modulus through a micro-mechanical approach. Precise control of the capillary number provides experimental data validating the theoretical description of the bubble as an equivalent elastic sphere. The good agreement between theoretical and measured elastic moduli confirms the generality of the study, which demonstrates that $\phi$ and $Ca$ entirely govern the overall response of the dispersions. More work remains to do in the limit of small capillary numbers ($Ca<1/4$), for which the predicted regime of increasing $\hat{G}(\phi, Ca)$ could not be experimentally investigated. Emulsions may not be the most relevant matrices for that study and dedicated experiments on even softer media could be more appropriate.

\bibliography{biblio}

\end{document}